\documentstyle[12pt,aasms4,psfig]{article}

\begin{document}

\title{Remnants of the Sagittarius Dwarf Spheroidal Galaxy around the
young globular cluster Palomar 12 }

\author{D. Mart\'\i nez-Delgado}
\affil{Instituto de Astrof\'\i sica de Canarias, E38200 La Laguna,
Tenerife, Canary Islands, Spain} \authoremail{ddelgado@ll.iac.es}

\author{R. Zinn}
\affil{Yale University, Department of Astronomy, P.O. Box 208101, New Haven, CT 06520-8101}
\authoremail{zinn@astro.yale.edu}

\author{R. Carrera}
\affil{Instituto de Astrof\'\i sica de Canarias,
E38200  La Laguna, Tenerife, Canary Islands, Spain}
\authoremail{rcarrera@ll.iac.es}

\author{C. Gallart}
\affil{Instituto de Astrof\'\i sica de Canarias 
E38200  La Laguna, Tenerife, Canary Islands, Spain}
\authoremail{carme@ll.iac.es}

\begin{abstract}

Photometry of a large field around the young globular cluster Palomar
12 has revealed the main-sequence of a low surface-brightness stellar
system.  This main-sequence is indicative of a stellar population that
varies significantly in metallicity and/or age, but in the mean is
more metal poor than Pal 12.  Under different assumptions for the
properties of this population, we find distances from the Sun in the range
17-24 kpc, which encompasses the distance to Pal 12, $19.0\pm0.9$ kpc.
The stellar system is also detected in a field 2$\arcdeg$ North of Pal
12, which indicates it has a minimum diameter of $\sim0.9$ kpc.  The orbit
of Pal 12 (Dinescu et al. 2000), the color-magnitude diagram of the
stellar system, their positions on the sky, and their distances
suggest that they are debris from the tidal disruption of the Sgr dSph
galaxy.  We discuss briefly the implications for the evolution of Sgr
and the Galactic halo.

\end{abstract}

\section{Introduction}

 The picture of
building the galactic halo from merging ``fragments'' resembling dwarf
galaxies, which Searle \& Zinn (1978, hereafter SZ)
proposed on the basis of the properties of the Milky Way globular
clusters, is often regarded as the local manifestation of the hierarchical
galaxy formation theory whereby dwarf galaxies were the first to form
and their subsequent mergers created large galaxies  (e.g. Moore et al.
1998; Navarro, Frenk \& White 1995). The Milky Way continues to be one of the best places to test this
picture, and much of the recent work in this area has focused on the
identification of stellar streams left behind in the halo after the
accretion of low-mass satellite galaxies (e.g. Yanny et al. 2000;
Ibata et al. 2001, Dohm-Palmer et al. 2001; Vivas et al. 2001).  The
prototypical case of accretion is the present-day tidal destruction of
the Sgr dwarf spheroidal (dSph) galaxy, which has deposited at least 4
and probably 5 (Dinescu et al. 2000) globular clusters in the halo and
produced a long stellar stream that wraps around the sky.  One of the
Sgr clusters, the very luminous cluster M54, and the most luminous
globular cluster in the Galaxy, $\omega$Cen, are suspected to be the
nucleus of Sgr (Sarajedini \& Layden 1995) and a now extinct dwarf
galaxy, respectively (Lee et al. 1999).  van den Bergh (2000) has
argued that the ``young'' globulars located in the outer halo might be
the nuclei of extinct dSph galaxies and has suggested that searches be
made for the vestiges of their parent systems.  We report here our
survey of a large field surrounding Palomar 12 (Pal 12)

Pal 12 is a remote globular cluster located at a distance of 19 kpc
from the Sun. It is among the youngest and most metal rich of the halo
globular clusters, which has fueled speculation that has been accreted
from another galaxy (e.g., Zinn 1993).  On the basis of its age and
radial velocity, Lin \& Richer (1992) argued that Pal 12 may have been
captured from the Magellanic Clouds.  The measurement of its proper
motion and the determination of its orbit around the Milky Way by
Dinescu et al. (2000) have shown that Pal 12 probably did not
originate in the Clouds but in the Sgr dSph galaxy (as suggested first by Irwin 1999). In this paper, we report the detection of a very low density stellar
system in the same direction and, to within the errors, the same
distance as Pal 12.  

\section{OBSERVATIONS AND DATA REDUCTION}

Pal 12 was observed in $B$ and $R$ Johnson--Cousins filters with the
Wide Field Camera (WFC) at the prime focus of the 2.5 m Isaac Newton
Telescope (INT) at the Roque de los Muchachos Observatory in June 2001.
 The WFC holds
four 4096 $\times$ 2048 pixel EEV CCDs with pixel size $0\farcs33$,
which provides a total field of about $35\arcmin \times
35\arcmin$. The field was centered at 10$\arcmin$ South of the center
of Pal 12 with the purpose of including the extra-tidal regions around
the cluster (Field 1). In addition, a field situated 2$\arcdeg$ North
of the cluster's center was also taken with the purpose of gauging the
field-star contamination (Field 2). However, this field is not far
enough away to avoid the large stellar system that we have discovered
in the environs of Pal 12. Another control field at Northern galactic
latitute, which was observed during the same run, has been used
instead. 

Bias and flatfield corrections were made with IRAF. DAOPHOT and
ALLSTAR (Stetson 1994) were then used to obtain the instrumental
photometry of the stars.  For the final photometric list, we selected
stars with $\sigma < 0.20$, $-1 < SHARP < 1$ and $0< CHI < 2$ as
provide by ALLSTAR. These criteria reject extended objects, so the
background contamination is expected to be only stellar-shaped
objects. The atmospheric extinction and the transformations to the
standard Johnson--Cousins photometric system were obtained from 52
measurements of 20 standard stars from the Landolt (1992) list. The
photometric conditions during the observing run were stable and
produced very small zero point errors for the photometric
transformation: $\pm$ 0.020 mag in $B$ and $\pm$ 0.016 in $R$. The offsets between the different WFC chips were obtained
from observations of several standard fields in each chip, and then
removed with a precision of better than 0.01 mag. Taking all
uncertainties into account, we estimate the total zero-point errors of
the photometry are $\sim0.025$ in both filters.

\section{THE LOW SURFACE BRIGHTNESS SYSTEM}

Figure 1a shows the color-magnitude diagram (CMD) for the WFC field
centered on Pal 12 (Field 1). In addition to the main-sequence (MS) of
the cluster, a MS-like feature is observed at bluer color, overlapping
the MS of the cluster at $B-R \sim 0.8$ and $V> 21.0$. This feature is
also clearly seen in the CMD of the field situated $\sim 2\arcdeg$
North of the center of Pal 12 (Field 2).

This unexpected feature of the Pal 12 diagram is more evident in
Figure 1b, which shows the CMD of the extra-tidal field of Pal
12. This includes the part of Field 1 beyond the tidal radius of the
cluster (i.e. for $r> 17\arcmin$) and one-half of the area of Field 2.
These regions have almost indistinguishable CMDs, with the only
difference that the density of stars in the MS feature is larger
in the cluster's field (Field 1).  The CMD of the control field
situated at (l,b)=$(28.7, 42.2)$ in the North is shown in Figure
1c. The total area of this field is the same as that plotted in
Fig. 1b. The absence of the MS feature in the control field is the
most striking difference between these Northern and Southern
hemisphere regions. The control field has slightly smaller l and $\mid b\mid$
than the fields in the direction to Pal 12.  If the halo is symmetric
about the galactic plane and has smooth density contours, either
spherical or flattened towards the plane, the control field and not
the fields in the direction to Pal 12 should have a higher density
of halo stars.  This is clearly not the case (compare the regions
bordered by $22>''V''>18$ \& $0.6<B-R<2.0$ in fig. 1b and 1c).

To check the significance of this detection, we constructed luminosity
functions of the stars in the ``V'' range 18.0-22.4, where the
photometric errors are $\le$ 0.05 and $\le$ 0.03 for the field near Pal 12
and the control field, respectively. The
color range was restricted to $0.6\le$B-R$\le1.1$, which should
encompass the MS and the subgiant branch of an old stellar system.
Its large width ensures that this comparison is insensitive to
reddening differences, which are expected to be very small
($\Delta$E(B-V)$\simeq0$, Schlegel et al. 1998, SFD).  Figure 2 shows
that for ``V''$<20.6$ there is little difference between the field
near Pal 12 and the control field.  Hovever, for
$20.8\le$``V''$\le22.4$, there is an excess of stars in the field near
Pal 12 over the control field, and for every 0.2 mag. bin in this
interval, this excess corresponds to several standard deviations.

Figure 2 also compares the luminosity functions with that of a model
stellar population, which was constructed from the isochrone
calculations of Girardi et al. (2000) for a metal abundance of Z=0.001
([Fe/H] =-1.3) and an age of 12.6 Gyr.  We imposed the same color
boundaries as above and transformed to apparent ``V'' magnitude by
adopting the same distance modulus and interstellar extinction as Pal
12.  To model the field stars, we fit a quadratic equation to the
luminosity function of the stars in the control field.  The solid line
in Figure 2 is the sum of this function plus the one for the 12.6 Gyr
stellar population.  It is normalized so that it matches the observed
number of stars in the field near Pal 12 for ``V''$ < 19.0$.  The
number of stars in the 12.6 Gyr system was varied until a rough match
was obtained between the model function and the observed points for
``V''$ > 21.5$.  The similarity between this model luminosity function
and the observed one over the whole range of ``V'' suggests that the
excess of stars over the control field is caused by a real stellar
system at approximately the same distance as Pal 12.  Unfortunately,
the number of stars is too small to tightly constrain the luminosity
function, and firm conclusions about the age or composition of the
system cannot be drawn from this match.

To estimate the surface brightness of this system, we used the control
field for an estimate of the field star contamination and produced a
decontaminated CMD following the procedure of Gallart, Aparicio \&
V\'\i lchez (1996).  Over the range $20.5\le$``V''$\le22.9$ and within
the same color range as above, there is an excess of $293\pm15$ MS
stars over the background.  The comparison of this number with that
found in a decontaminated CMD of the main body of the Sgr dSph (
Mart\'\i nez-Delgado et al. 2002) yields an estimate of $\Sigma$=
30.5$\pm$ 0.2 mag arcsec$^{-2}$.  While undoubtedly this system has a
very low surface brightness (LSB), we caution that this value is a
rough estimate and also note that there is probably a gradient in
$\Sigma$ across the field.  This value is the same found by 
Mateo et al. (1998) in their outermost detection of the Sgr stream, 
34$\arcdeg$ SE from the center of Sgr.

A comparison between Figures 1a and 1b shows that the extra-tidal
stellar population is bluer than that of Pal 12 and displays a
significant width in color, indicating the presence of a range of age
and/or metallicity and/or some depth along the line of sight, as
expected of a dwarf galaxy.  The simplest explanation is that this MS
feature is not due to Pal 12 but to a separate stellar population that
is part of a dSph galaxy or a tidal stream from one. The presence of
this feature throughout the area covered by Field 2 puts a lower limit
of 0.9 kpc on the extension of the system, assuming that it is at the
same distance as Pal 12 (19.0 kpc; see below). This size is compatible
with the width of a tidal stream crossing the field (e.g. Sgr tidal
stream; Newberg et al. 2002) or the projected tidal radius of a
typical dSph at the cluster's distance.\footnote{ A typical dSph with
a tidal radius of $\sim$ 1 kpc would have angular radius of 3 degrees
at this distance.}.

 While the comparison in Figure 2 between the
model and observed luminosity functions illustrates that the LSB system could be at the same distance as Pal 12, it is important to examine this question in
more detail.  In Figure 1b, the MS appears to terminate at
``V''=$20.48\pm0.10$.  The MS turnoff (i.e., the bluest point of the
MS) is probably fainter than this (e.g., the turnoff of the 12.6 Gyr
population in Fig. 2 occurs at 20.65, $M_{V}=4.14$ Girardi et
al. 2000), but this point cannot identified with certainty given the
small number of stars.  It seems unlikely that it could be fainter
than ``V''=20.9.  Adopting this value, a somewhat brighter turnoff
luminosity ($M_{V}=3.9$), and correcting for the interstellar
extinction (E(B-V)=$0.037\pm0.002$, SFD), we obtain 23.8 kpc for an upper
limit on the distance, if the LSB system is composed of a very old
stellar population.  For a lower limit under the same assumption, we
adopt 20.48 as the turnoff and a fainter turnoff luminosity
($M_{V}=4.2$), which yields a distance of 17.1 kpc.  It is likely that
the LSB population is a mixture of ages and compositions, which is
much harder to model.  Under the assumption that its stellar
population resembles the main body of the Sgr dSph galaxy, we can use
the decontaminated CMDs of the LSB system and Sgr (see above) to help
set the distance modulus of the LSB system.  These CMDs suggest that
the MS feature seen in Figure 1b is only the densest part of the MS
and that its ``termination'' at ``V''=20.48 corresponds to point in
Sgr where $M_{V}\approx3.7$ (Layden \& Sarajedini 2000).  This yields
a distance of 21.6 kpc for the LSB system, with an uncertainty of
$\sim2$ kpc.  The distance to Pal 12 is $19.0 \pm 0.9$ kpc (Rosenberg
et al. 1998, but using E(B-V) from SFD), which places it near the
middle of the range of estimates for the distance to the LSB system.
If they are unrelated, this is a very remarkable coincidence.

Dinescu et al. (2000) have shown on the basis of the orbit of Pal 12
that the Sgr dSph galaxy is likely to be its parent galaxy. Recent
surveys of Sgr (Mateo, Olszewski \& Morrison 1998; Ibata et al. 2001;
Majewski et al. 1999; Yanny et al. 2000; Mart\'\i nez-Delgado et
al. 2001; Mart\'\i nez-Delgado, G\'omez-Flechoso \& Aparicio 2002)
have shown that this galaxy forms a giant stream that wraps completely
around the Milky Way in an almost polar orbit, in good agreement with
the predictions of theoretical models (G\'omez-Flechoso, Fux \&
Martinet, 1999; Johnston, Sigurdsson \& Hernquist 1999 ; Helmi \&
White 2001).  The position of Pal 12 on the sky ($\sim 40\arcdeg$ from
the Sgr's main body) is very close to that predicted for the Sgr
Southern stream (see Mart\'\i nez-Delgado et al. 2001). The predicted
distance for the Sgr's tidal stream at the cluster's position is
$\sim19$ kpc (Mart\'\i nez-Delgado et al. 2002), in good agreement
with the distance obtained above for th LSB system. 

The CMD of the LSB system also resembles those of
previous detections of the Sgr Southern stream. Majewski et al. (1999)
reported a MS turnoff at V=21 in a field situated at (l,b)=
(11$\arcdeg$, -40$\arcdeg$), which is $\sim 20\arcdeg$ from Pal
12. In the Sloan Digitized Sky Survey (SDSS), Newberg et al. (2002)
found a similar structure (named S167-54-21.5) in the CMD of a long,
narrow region centered on the celestial equatorial and $15\arcdeg <
\alpha < 0\arcdeg$, that could be also part of the Sgr tidal stream.

As we noted above, the MS of the LSB system is bluer than the MS of
Pal 12 and extends to approximately the same bright magnitude. These
characteristics suggest that it is more metal poor than Pal 12
($[Fe/H]$=-1.0; Brown, Wallerstein \& Zucker 1997) and may be as old
or older than it, since lower metallicity isochrones have brighter
turnoffs for a given age.  The MS feature has intermediate color
between the $[Fe/H]$=-1.7 and -0.7 isochrones of Bertelli et
al. (1994), which yields a mean value of $\sim -1.2$. This is similar
to the metallicity of the metal-poor component of the main body of Sgr
($[Fe/H]$= -1.3; Layden \& Sarajedini 2000) and the giants in the
Northern stream of Sgr (Dohm-Palmer et al. 2001).

\section{ DISCUSSION }

Our detection of a LSB stellar system around Pal 12 strengthens
considerably the contention of Dinescu et al. (2000) that Pal 12
originated in the Sgr dSph galaxy.  The disruption of Sgr has
therefore released at least 5 globular clusters to the Galactic halo (Ibata, Gilmore \& Irwin 1994; Irwin 1999), and additional ones may remain undetected among the globular clusters
in the outer halo.  Two of the Sgr clusters (M54 and Ter 8) are
similar in age to the oldest globular clusters in the Milky Way.  A
third (Arp 2) is slightly younger than these other two (Buonanno et
al. 1998; Layden \& Sarajedini 2000).  The much younger Sgr cluster
(Ter 7) and Pal 12 (Buonanno et al. 1998; Rosenberg et al. 1998) are
several Gyrs younger than the other three Sgr clusters, and they are
the youngest and most metal-rich globular clusters in the galactic
halo.  Sgr resembles the Fornax dSph galaxy in having 5 globular
clusters, although their histories of cluster formation are
significantly different.  Only one of the Fornax clusters is clearly
younger than the other four, and this cluster (cluster 4, Buonanno et
al. 1999) is both older and less metal rich than the youngest Sgr
clusters (Ter 7 and Pal 12).

Spectroscopic observations by Brown et al. (1997) have shown that Pal
12 is almost unique among halo globular clusters by having
$[\alpha/Fe]$=0.0 instead of $\approx0.3$.  This relatively low $\alpha$
abundance is a sign of metal enrichment by Type Ia supernovae (Brown
et al. 1997).  The measurement of the abundance ratios in
several red giants in Sgr by Smecker-Hane \& McWilliam (1999) have
shown that two metal-poor stars ($[Fe/H]<-1$) are $\alpha$ enhanced
while 9 others with $[Fe/H]>-1$ have $[\alpha/Fe]\sim0.0$.  The
chemical composition of Pal 12 ($[Fe/H]=-1$, $[\alpha/Fe]$=0.0, Brown
et al. 1997) is also consistent with membership in Sgr.

The three ``young'' Sgr clusters (Arp 2, Ter 7, Pal 12) constitute one
quarter of the ``young'' globular clusters in the Galactic halo (van
den Bergh 2000).  Many of the other young halo clusters (perhaps all),
and at least some of the old clusters may have originated in
dwarf galaxies that later merged with the Milky Way.  The stellar
streams from Sgr show that this process not only added globular
clusters to the halo, but also contributed stars of different ages and
metallicities, as predicted by the SZ scenario.

We are grateful to M. A. G\'omez-Flechoso, for making her Sgr model
available to us and for several fruitful suggestions. We also thanks
to R. Garc\'\i a-L\'opez, Steve Majewski and M. Shetrone for many
useful comments. This work is based on observations made with the 2.5
m INT operated  by the
Isaac Newton Group in the Spanish Observatorio del Roque de Los
Muchachos of the Instituto de Astrof\'\i sica de Canarias.  RZ was
supported by grant AST 00-98428 from the National Science Foundation.

\newpage

\begin{deluxetable}{lcccccc}
\tablenum{1}
\tablewidth{400pt}
\tablecaption{Positions and Integration Times
\label{Target_fields}}
\tablehead{
\colhead{Field} & \colhead{R.A.(J2000.0)} & \colhead{Dec} & \colhead{l} &
\colhead{b} & \colhead{$t_{B}$(s)}& \colhead{$t_{R}$(s)}}
\startdata
Field 1    & $21^{h} 46^{m} 38\fs9$ & $-21\arcdeg 25\arcmin 00\arcsec$ &  30.3 & -47.7 & 3600 & 1800 \nl
Field 2  & $21^{h} 49^{m} 21\fs7$ & $-19\arcdeg 20\arcmin 10\arcsec$ & 33.5 & -47.7    & 3900 & 3600\nl
Control    & $16^{h} 11^{m} 04\fs0$ & $+14\arcdeg 57\arcmin 29\arcsec$ & 28.7 & 42.2  & 9200 & 9000 \nl
\enddata
\end{deluxetable}

\newpage
Figure 1. {\it panel a)} Color-magnitude diagram for the WFC field
centered on Pal 12 (Field 1). The narrow main-sequence (MS) of the
cluster is clearly delineated.  Notice that it is surrounded by a
sparser and wider MS-like feature at bluer color; {\it panel b)}
displays the CMD for an extra-tidal field of $35\arcmin \times
35\arcmin$, half of which is from Field 1 but outside of the cluster's
tidal radius ( $r> 17\arcmin$) and the other half is from Field 2,
situated $\sim 2\arcdeg$ North of Pal 12.  The MS feature is most
clearly observed at $B-R \sim 0.8$ and $21 <V<23 $; {\it panel c)}
shows the CMD of our control field, which is the same size as the
field shown in 1b.

Fig. 2. Comparison of the luminosity functions of the extra-tidal
field of Pal 12 and the control field. The solid line is the model LF
for a stellar population with age 12.6 Gyr and Z=0.001, adopting the
distance and reddening of Pal 12. The error bars are $\pm\sqrt{N}$,
where $N$ is the number of stars in each bin.

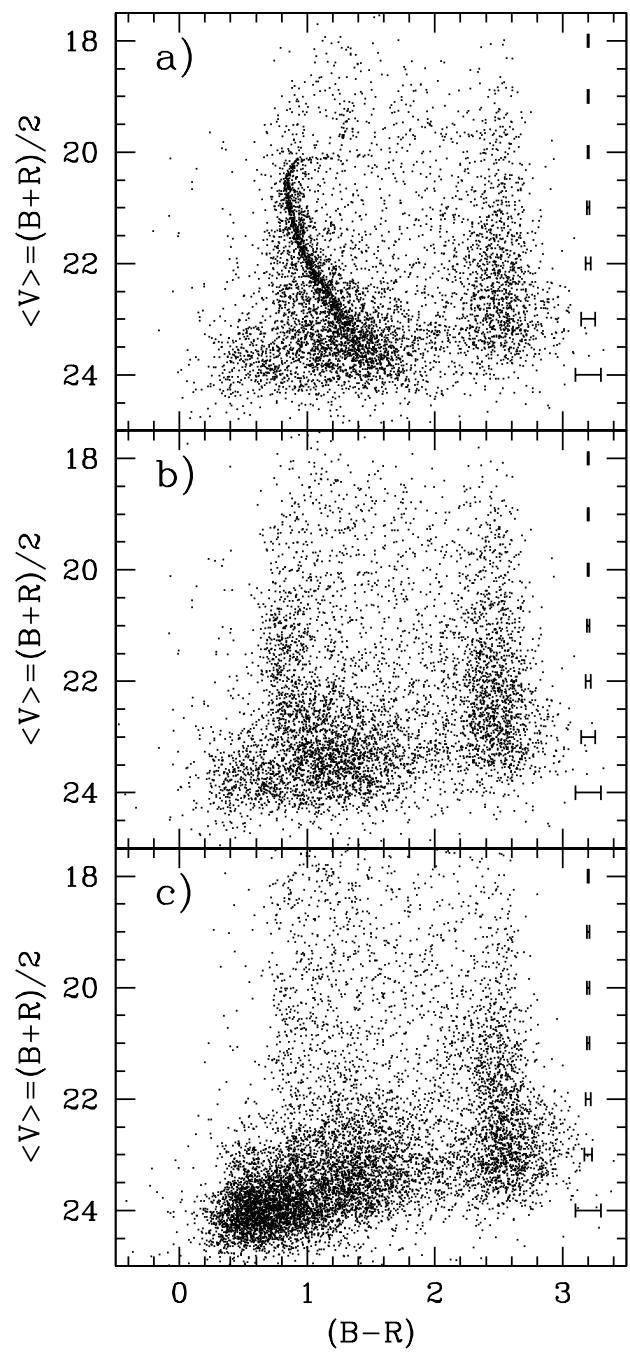
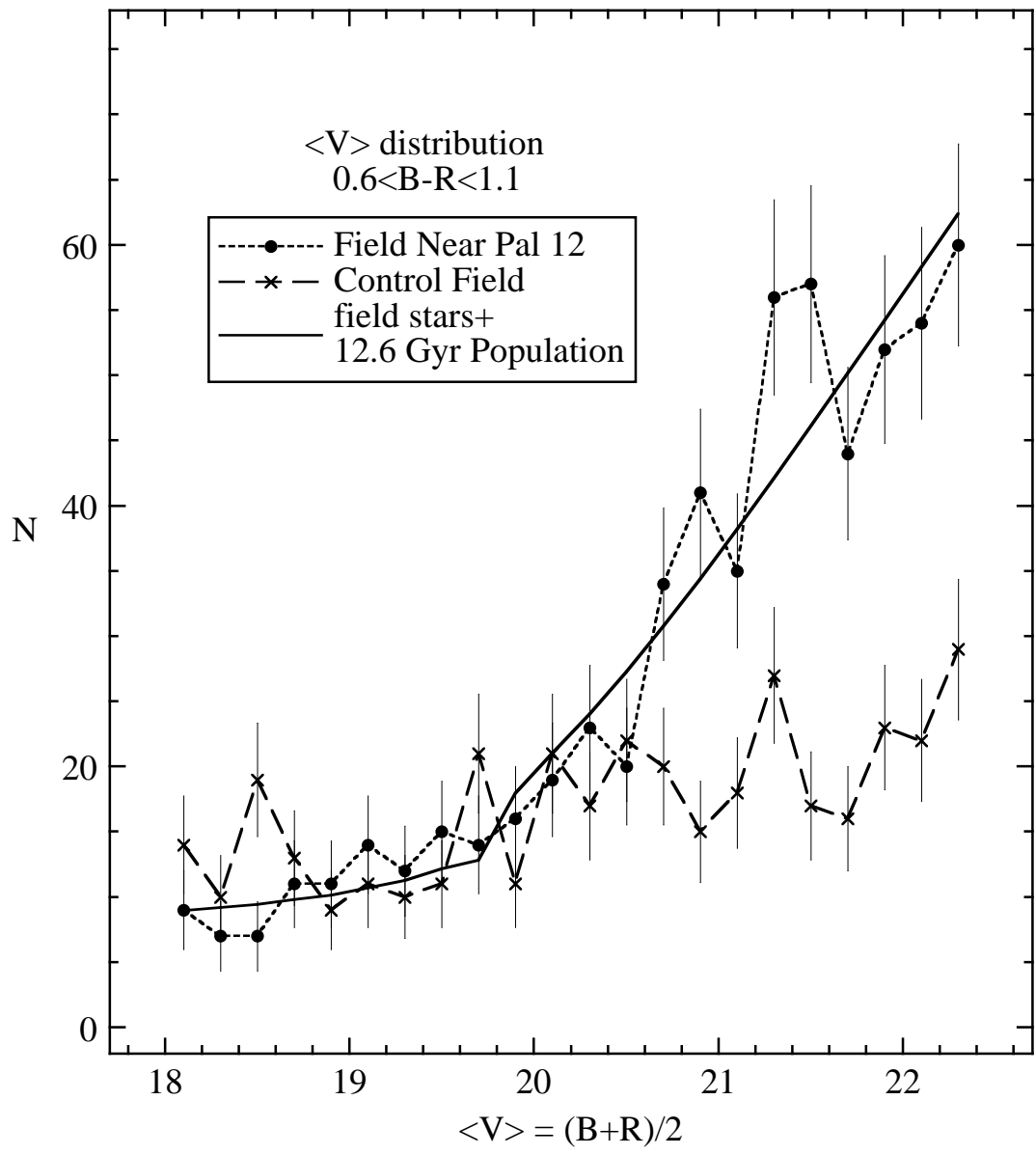
\end{document}